\documentclass[twocolumn,superscriptaddress,letter]{revtex4}%
\usepackage{amssymb}
\usepackage{color}
\usepackage{graphicx}
\usepackage{dcolumn}
\usepackage{bm}
\usepackage[header,title,page,titletoc]{appendix}
\usepackage{amsmath}
\usepackage{amsfonts}%
\providecommand{\U}[1]{\protect \rule{.1in}{.1in}}

\begin{document}
\title{Mott Insulator-Superfluid Transition in a Generalized Bose-Hubbard Model with
Topologically Non-trivial Flat-Band}
\author{Xing-Hai Zhang}
\affiliation{Department of Physics, Beijing Normal University, Beijing 100875, People's
Republic of China}
\author{Su-Peng Kou}
\email{spkou@bnu.edu.cn}
\affiliation{Department of Physics, Beijing Normal University, Beijing 100875, People's
Republic of China}

\begin{abstract}
In this paper, we studied a generalized Bose-Hubbard model on a
checkerboard lattice with topologically nontrivial flat-band. We
used mean-field method to decouple the model Hamiltonian and
obtained phase diagram by Landau theory of second-order phase
transition. We further calculate the energy gap and the dispersion
of quasi-particle or quasi-hole in Mott insulator state and found
that in strong interaction limit the quasi-particles or the
quasi-holes also have flat bands.

\end{abstract}
\maketitle

\section{INTRODUCTION}

The Mott insulator (MI) -- superfluid transition (SF) of ultracold bosons in
optical lattices\cite{Jaksch, Greiner} has become a hot topic in quantum
simulation\cite{Buluta} and a great deal of works have been done. The Bose-Hubbard model proposed in Ref. \cite{Fisher} have been widely investigated both theoretically and experimentally so that a comprehensive understanding of this many-body system can be achieved.

On the other hand, the states with topological order become another hot topic and have been explored from different aspects.
Since first discovered in two-dimensional electron gas with Landau levels in
strong magnetic field, integer and fractional quantum Hall effect (IQHE and
FQHE) have attracted great attention. Haldane pointed out that IQHE may appear
in honeycomb lattice without Landau levels\cite{haldane}. However, FQHE in a
lattice model without Landau levels had not been discovered until the proposal
of the lattice models with topologically nontrivial flat-band based on the
mechanism of quadratic band touching\cite{Tang, Sun, Neupert}. In these
lattice models (kag\'{o}me lattice, $\pi$-flux square lattice and honeycomb
lattice), the hoppings of next-nearest or next-next-nearest (NNN and NNNN)
neighbor besides nearest neighbor(NN) are introduced to achieve the flat-bands
with a high flatness ratio (the flatness ratio defined as the band gap over the band width) in
different lattice models. To obtain FQHE states, people always consider these
tight-binding models with the NN and
NNN repulsions for fermions and corresponding
tight-binding models under a hard-core condition for bosons. Then the
numerical results confirm the existence of topologically nontrivial flat-band.
For example, in Ref.(\cite{Wang}), in an extended bosonic Haldane model with
the NN and NNN repulsions under
a hard-core condition, the $\nu=1/2$ and the $\nu=1/4$ bosonic FQHE states are predicted.

So we may ask questions : in a bosonic lattice model of
topologically nontrivial flat-band with on-site Coulomb interaction,
what's the ground state and whether the topologically nontrivial
flat-band affects the superfluid state and Mott state? This becomes
the starting point of this paper. In this paper, we study the MI-SF
transition in a bosonic system of the checkerboard model with
topological flat-band\cite{Sun}, of which there is no NN or NNN
repulsions. Instead, we replace the hard-core condition by a tunable
on-site repulsive interaction and consider a generalized
Bose-Hubbard model with the hopping parameters as that of the models
with topologically nontrivial flat-band. We will use the mean field
approach to obtain the quantum phase transitions between MI phase
and SF phase\cite{Oosten, Lim}. Then the elementary excitations are
studied in Mott insulator state. We derive the energy gap and the
dispersion of quasi-particle or quasi-hole.

The paper is organized as follow: We first introduce the generalized
Bose-Hubbard model with the hopping parameters as that of the models with
topologically nontrivial flat-band in Sec. II, then study this model in Sec.
III by mean field approach. And in Sec. III we obtain the phase diagrams of
this model. The properties of collective modes are analyzed in the last
subsection in Sec. IV. Finally we draw the conclusion in Sec. V.

\section{The model Hamiltonian}

\begin{figure}[pbh]
\begin{center}
\includegraphics[width=0.25\textwidth]{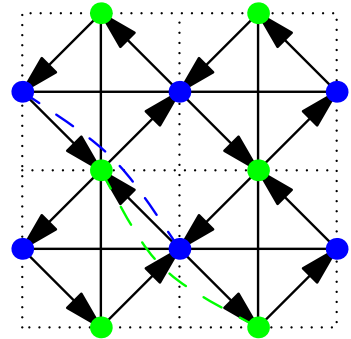}
\end{center}
\caption{The scheme of the square lattice, of which the two different lattice sites in blue or green colors are two sublattices A and B. The NN hoppings $t$ are represented by lines with arrows
whose directions are sign of the phase $\varphi$. The dotted lines represent the NNN hoppings which are $-t'$ between green sites(sublattice A) and $t'$ between blue sites(sublattice B). Two dashed lines representing the NNNN hoppings are displayed in the scheme, too.}%
\label{fig:fig1}%
\end{figure}

We first generalize the fermionic topological flat-band model to bosonic
system with the Hamiltonian
\begin{align}
H &  =-t\sum_{\langle j,l\rangle}\left[  b_{\mathbf{r}_{j}}^{\dagger
}b_{\mathbf{r}_{l}}\exp \left(  i\varphi_{jl}\right)  \right]  \pm t^{\prime
}\sum_{\langle \langle j,l\rangle \rangle}b_{\mathbf{r}_{j}}^{\dagger
}b_{\mathbf{r}_{l}}\nonumber \\
&  -t^{\prime \prime}\sum_{\langle \langle \langle j,l\rangle \rangle \rangle
}b_{\mathbf{r}_{j}}^{\dagger}b_{\mathbf{r}_{l}}+\frac{U}{2}\sum_{j}%
n_{\mathbf{r}_{j}}(n_{\mathbf{r}_{j}}-1)\label{eq:eq1}\\
&  -\mu \sum_{j}b_{\mathbf{r}_{j}}^{\dagger}b_{\mathbf{r}_{j}}+\text{H.c.}%
\nonumber
\end{align}
where $b_{\mathbf{r}_{j}}^{\dagger}$ ($b_{\mathbf{r}_{j}}$) is the bosonic
creation (annihilation) operator of the $j$-th site with position vector
$\mathbf{r}_{j}$. $\langle j,l\rangle$, $\langle \langle j,l\rangle \rangle$ and
$\langle \langle \langle j,l\rangle \rangle \rangle$ denote that $j$ and $l$ are
the NN, NNN and NNN sites, respectively. The phase factor of NN hopping
$\varphi_{jl}$ equals $\pm \varphi$ (Figure \ref{fig:fig1}). $t$, $t^{\prime},$
$t^{\prime \prime}$ are the hopping parameters for NN hoppings, NNN hoppings,
NNNN hoppings, respectively. $U$ is the on-site interaction strength and $\mu$
is the chemical potential.

In the following parts we will use mean field approach to calculate above
bosonic topological flat-band model. In addition, to show the effect of the
flat-band on MI-SF transition, we also calculate the generalized Bose-Hubbard
model without topological flat-band and compare their phase diagram.

\section{Mean-field approximation and phase diagram}

In this section we derive the decoupled effective Hamiltonian from
Eq.(\ref{eq:eq1}) by mean-field approximation and calculate the corresponding
perturbation energy up to second order. Then the equation of critical line of
MI-SF transition is obtained by Landau theory of phase transition.

\subsection{Mean-field approximation}

In the strong-coupling Limit ($t\ll U$), a localized superfluid order
parameter $\psi_{\mathbf{r}_j}=\langle b_{\mathbf{r}_{j}}^{\dagger}\rangle=\langle
b_{\mathbf{r}_{j}}\rangle$ is introduced so as to rewrite the hopping terms by
a consistent mean-field method as
\begin{equation}
b_{\mathbf{r}_{j}}^{\dagger}b_{\mathbf{r}_{l}}\simeq b_{\mathbf{r}_{j}%
}^{\dagger}\psi_{\mathbf{r}_l}+b_{\mathbf{r}_{l}}\psi_{\mathbf{r}_j}-\psi_{\mathbf{r}_j}\psi_{\mathbf{r}_l}.\label{eq:eq2}%
\end{equation}
Thus the Hamiltonian can be written as
\begin{align}
H^{\text{eff}}\  &  =-t\cos \varphi \sum_{j}\sum_{l=1}^{z}\left[  b_{\mathbf{r}%
_{j}}^{\dagger}+b_{\mathbf{r}_{j}}-\psi_{\mathbf{r}_{j}}\right]
\psi_{\mathbf{r}_{j}+\boldsymbol{\tau}_{l}}\nonumber \\
&  \pm t^{\prime}\sum_{j}\sum_{l=1}^{z^{\prime}}\left[  b_{\mathbf{r}_{j}%
}^{\dagger}+b_{\mathbf{r}_{j}}-\psi_{\mathbf{r}_{j}}\right]  \psi
_{\mathbf{r}_{j}+\boldsymbol{\tau}_{l}^{\prime}}\nonumber \\
&  -t^{\prime \prime}\sum_{j}\sum_{l=1}^{z^{\prime \prime}}\left[
b_{\mathbf{r}_{j}}^{\dagger}+b_{\mathbf{r}_{j}}-\psi_{\mathbf{r}_{j}}\right]
\psi_{\mathbf{r}_{j}+\boldsymbol{\tau}_{l}^{\prime \prime}}\nonumber \\
&  +\frac{U}{2}\sum_{j}n_{\mathbf{r}_{j}}(n_{\mathbf{r}_{j}}-1)-\mu\sum
_{j}b_{\mathbf{r}_{j}}^{\dagger}b_{\mathbf{r}_{j}}\ ,\label{eq:eq3}%
\end{align}
where $z,$ $z^{\prime},$ $z^{\prime \prime}$ denote the coordination numbers of
NN, NNN and NNNN hoppings and $\boldsymbol{\tau}_{l},$ $\boldsymbol{\tau}%
_{l}^{\prime},$ $\boldsymbol{\tau}_{l}^{\prime \prime}$ are the corresponding
translation vectors, respectively. Hence the decoupled effective Hamiltonian can be
rewritten with respect to $\mathbf{r}_j$ as
\begin{align}
H_{\mathbf{r}_j}^{\text{eff}} &  =-zt\cos \varphi(b_{\mathbf{r}_j}^{\dagger
}+b_{\mathbf{r}_j}-\psi_{\mathbf{r}_j})\psi_{\mathbf{r}_j+\boldsymbol{\tau}%
}\nonumber \\
&  -zt^{\prime \prime}(b_{\mathbf{r}_j}^{\dagger}+b_{\mathbf{r}_j}-\psi
_{\mathbf{r}_j})\psi_{\mathbf{r}_j+\boldsymbol{\tau}^{\prime \prime}}\nonumber \\
&  +\frac{U}{2}n_{\mathbf{r}_j}(n_{\mathbf{r}_j}-1)-\mu n_{\mathbf{r}_j%
}\ .\label{eq:eq4}%
\end{align}

Since there are two sublattices of this lattice (A and B), we can define two
order parameters $\psi_{A}=\langle b_{\mathbf{r}_{j\in A}}^{\dagger}\rangle$
and $\psi_{B}=\langle b_{\mathbf{r}_{j\in B}}^{\dagger}\rangle$ that
correspond to the Bose condensations on A site and on B site, respectively. The
order parameters satisfy the condition $\psi_{A}=\psi_{\mathbf{r}_{j}}%
=\psi_{\mathbf{r}_{j}+\boldsymbol{\tau}^{\prime}}=\psi_{\mathbf{r}%
_{j}+\boldsymbol{\tau}^{\prime \prime}}$ and $\psi_{B}=\psi_{\mathbf{r}%
_{j}+\boldsymbol{\tau}}$ if site $j$ denotes an A site and similar condition
for B site can be obtained if we substitute A with B. Then we can construct an
effective Hamiltonian of a two-site cell with the condition $z=z^{\prime
}=z^{\prime \prime}$ as
\begin{align}
  H^{\text{eff}}  &  =(zt)^{-1}(H_{1}+H_{2})=(zt)^{-1}(H^\text{eff}_{\mathbf{r}_j%
  }+H^\text{eff}_{\mathbf{r}_j+\boldsymbol{\tau}})\nonumber \\
&  =\frac{\overline{U}}{2}\left(  n_{1}^{2}+n_{2}^{2}-n\right)  -\bar{\mu
}n+2\psi_{1}\psi_{2}\cos \varphi \nonumber \\
&  +\bar{t}^{\prime \prime}\psi_{1}^{2}+\bar{t}^{\prime \prime}\psi_{2}%
^{2}+(\psi_{2}\cos \varphi+\bar{t}^{\prime \prime}\psi_{1})V_{1}\nonumber \\
&  +(\psi_{1}\cos \varphi+\bar{t}^{\prime \prime}\psi_{2})V_{2}\ ,
\label{eq:eq5}%
\end{align}
where $\psi_{1}=\psi_{A}$, $\psi_{2}=\psi_{B}$ and $\overline{U}=\frac{U}{zt}%
$, $\bar{\mu}=\frac{\mu}{zt}$, $\bar{t}^{\prime \prime}=\frac{t^{\prime \prime}%
}{zt}$, $V_{i}=-(b_{i}^{\dagger}+b_{i})$. We then divide the effective
Hamiltonian into two terms: the unperturbed term
\begin{equation}
H^{(0)}=\frac{\overline{U}}{2}(n_{1}^{2}+n_{2}^{2}-n)-\bar{\mu}n+2\psi_{1}%
\psi_{2}\cos \varphi+\bar{t}^{\prime \prime}\psi_{1}^{2}+\bar{t}^{\prime \prime
}\psi_{2}^{2}\ , \label{eq:h0}%
\end{equation}
and the perturbation term
\begin{equation}
H^{\prime}=(\psi_{1}\bar{t}^{\prime \prime}+\psi_{2}\cos \varphi)V_{1}+(\psi
_{1}\cos \varphi+\psi_{2}\bar{t}^{\prime \prime})V_{2}\ . \label{eq:h'}%
\end{equation}
Hence the unperturbed ground energy is given by
\begin{equation}
E^{(0)}=\frac{\overline{U}}{2}(n_{1}^{2}+n_{2}^{2}-n)-\bar{\mu}n+2\psi_{1}%
\psi_{2}\cos \varphi+\bar{t}^{\prime \prime}\psi_{1}^{2}+\bar{t}^{\prime \prime
}\psi_{2}^{2}\ . \label{eq:E0}%
\end{equation}

As first-order perturbation of energy vanishes, we get second-order energy perturbation
\begin{align}
E^{(2)}  &  =-\frac{(\psi_{2}\cos \varphi+\bar{t}^{\prime \prime}\psi_{1}%
)^{2}(\overline{U}+\bar{\mu})}{(-n_{1}\overline{U}+\bar{\mu})\left[
(n_{1}-1)\overline{U}-\bar{\mu}\right]  }\nonumber \\
&  -\frac{(\psi_{1}\cos \varphi+\bar{t}^{\prime \prime}\psi_{2})^{2}%
(\overline{U}+\bar{\mu})}{(-n_{2}\overline{U}+\bar{\mu})\left[  (n_{2}%
-1)\overline{U}-\bar{\mu}\right]  } \ . \label{eq:eq7}%
\end{align}
Thus energy up to second-order is obtained as
\begin{equation}
E(\psi_{1},\psi_{2})\approx a_{0}+a_{2}\psi_{1}^{2}+c_{2}\psi_{1}\psi
_{2}+b_{2}\psi_{2}^{2}+\mathcal{O}(\psi^{2})\ , \label{eq:eq8}%
\end{equation}
where
\begin{align}
a_{0}    =&\frac{\overline{U}}{2}(n_{1}^{2}+n_{2}^{2}-n)-\bar{\mu}n\nonumber \\
a_{2}    =&\bar{t}^{\prime \prime}-\frac{\bar{t}^{\prime \prime2}(\overline
{U}+\bar{\mu})}{(-n_{1}\overline{U}+\bar{\mu})\left[  (n_{1}-1)\overline
{U}-\bar{\mu}\right]  }\nonumber \\
  &-\frac{(\overline{U}+\bar{\mu})\cos^{2}\varphi}{(-n_{2}\overline{U}%
+\bar{\mu})\left[  (n_{2}-1)\overline{U}-\bar{\mu}\right]  }\nonumber \\
b_{2}    =&\bar{t}^{\prime \prime}-\frac{(\overline{U}+\bar{\mu})\cos
^{2}\varphi}{(-n_{1}\overline{U}+\bar{\mu})\left[  (n_{1}-1)\overline{U}%
-\bar{\mu}\right]  }\nonumber \\
&  -\frac{\bar{t}^{\prime \prime2}(\overline{U}+\bar{\mu})}{(-n_{2}\overline
{U}+\bar{\mu})\left[  (n_{2}-1)\overline{U}-\bar{\mu}\right]  }\nonumber \\
c_{2}    =&2\cos \varphi-\frac{2\bar{t}^{\prime \prime}(\overline{U}+\bar{\mu
})\cos \varphi}{(-n_{1}\overline{U}+\bar{\mu})\left[  (n_{1}-1)\overline
{U}-\bar{\mu}\right]  }\nonumber \\
&  -\frac{2\bar{t}^{\prime \prime}(\overline{U}+\bar{\mu})\cos \varphi}%
{(-n_{2}\overline{U}+\bar{\mu})\left[  (n_{2}-1)\overline{U}-\bar{\mu}\right]
}\ . \label{eq:eq9}%
\end{align}

\subsection{Phase diagram}

The MI-SF phase transition occurs on the condition when Gaussian curvature is
zero at the point $\psi_{1}=0,\psi_{2}=0$, namely
\begin{equation}
\frac{\partial^{2}E}{\partial \psi_{1}^{2}}\bigg |_{(0,0)}%
\frac{\partial^{2}E}{\partial \psi_{2}^{2}}\bigg|_{(0,0)}-\left(
\frac{\partial^{2}E}{\partial \psi_{1}\partial \psi_{2}%
}\bigg|_{(0,0)}\right)  ^{2}=0 \ .
\end{equation}
It implies that
\begin{equation}
4a_{2}b_{2}-c_{2}^{2}=0\ . \label{eq:eq16}%
\end{equation}
With $n_{1}=n_{2}=n$, considering the symmetry between two sublattices,
$a_{2}=b_{2}$, we have $2a_{2}=\pm c_{2}$. Solve the equation for $\bar{\mu}$,
we can get the boundary condition
\begin{align}
\bar{\mu} =  &  \frac{1}{2}\Bigg[(2n-1)\overline{U}-(\bar{t}^{\prime \prime
}-\cos \varphi)\nonumber \\
&  \pm \sqrt{\overline{U}^{2}-2(2n+1)(\bar{t}^{\prime \prime}-\cos
\varphi)\overline{U}+(\bar{t}^{\prime \prime}-\cos \varphi)^{2}}\Bigg]\ .
\label{eq:mu_bound2}%
\end{align}
for negative $\overline{U}$, namely positive $U$ as $t=-1$.

\begin{figure}[ptbh]
\begin{center}
\includegraphics[width=0.5\textwidth]{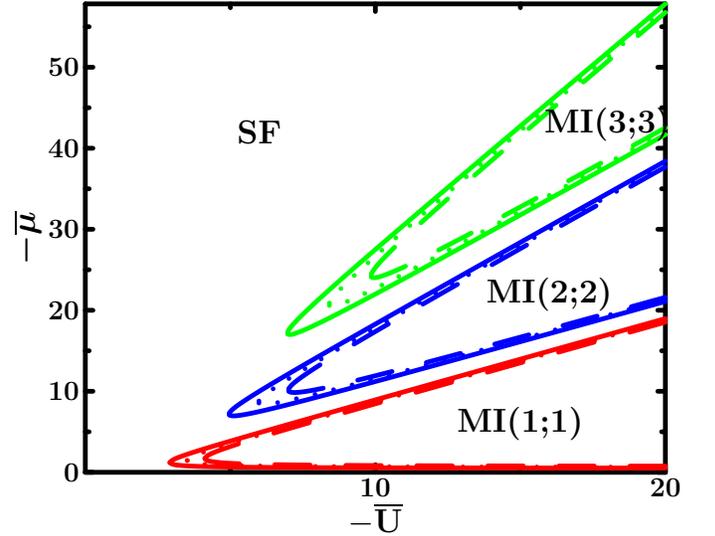}
\end{center}
\caption{The phase diagram of MI-SF transition. The solid lines are the phase
boundaries of topological non-trivial flat-band case with $t=-1$, $\bar
{t}^{\prime}=-\frac{1}{2+\sqrt{2}}$, $\bar{t}^{\prime \prime}=\frac{1}%
{2+2\sqrt{2}}$ and $\varphi=\frac{\pi}{4}$. The dashed lines are the phase
boundaries of the case with $t=-1$, $\bar{t}^{\prime}=0$, $\bar{t}^{\prime \prime}=0$ and
$\varphi=\frac{\pi}{4}$. The dotted lines are phase boundaries of the case with $t=-1$,
$\bar{t}^{\prime}=-\frac{1}{2(2+\sqrt{2})}$, $\bar{t}^{\prime \prime}=\frac
{1}{2(2+2\sqrt{2})}$ and $\varphi=\frac{\pi}{4}$. On the other hand, the red lines represent phase boundaries of
configuration $\{1;1\}$, blue ones $\{2;2\}$, and green ones $\{3;3\}$.}%
\label{fig:phase-diagram}%
\end{figure}

The phase diagram is displayed in Figure \ref{fig:phase-diagram}. In Figure
\ref{fig:phase-diagram}, there are three MI-SF phase transition lines for each
particle number configuration. The particle number configuration
$\{n_{1};n_{2}\}$ represents that there are $n_{1}$ bosons on each A lattice
site while $n_{2}$\ bosons on each B\ lattice site. Due to the existence of
inversion symmetry, we have $n_{1}=n_{2}$.

The solid lines are phase boundaries of the flat-band case with $t=-1$, $\bar
{t}^{\prime}=-\frac{1}{2+\sqrt{2}}$, $\bar{t}^{\prime \prime}=\frac{1}%
{2+2\sqrt{2}}$ and $\varphi=\frac{\pi}{4}$. We can see that quantum
phase transition occurs when $\overline{U}$ increases by fixing
$\bar{\mu}$. The phase diagram is similar to the traditional
Bose-Hubbard model with only nearest hopping term\cite{Oosten, Lim}.
To show the effect of flat-band, we also calculate other two cases
without topological flat-bands: the $\varphi $-flux Bose-Hubbard
model with only nearest hopping term or $\bar{t}^{\prime
}={t}^{\prime \prime}=0,$ $\varphi=\frac{\pi}{4}$ (the dashed lines)
and the generalized Bose-Hubbard model with smaller NNN and NNNN
hopping parameters compared with flat-band case, namely
$\varphi=\frac{\pi}{4}$ and $\bar
{t}^{\prime}=-\frac{1}{2(2+\sqrt{2})}$, $\bar{t}^{\prime
\prime}=\frac {1}{(2+2\sqrt{2})}$ (the dotted lines). From these
results, we can see that, the superfluid phase of flat-band model
becomes larger compared with the $\varphi$-flux Bose-Hubbard model
with only nearest hopping term.

\section{Collective modes in Mott phase}

In this section, we first derive the effective action of this bosonic checkerboard
model of flat-band. Then we calculate the dispersion and the energy
gap of excitations in Mott phase.

\subsection{The effective action}

With the complex functions $b_{i}^{\ast}(\tau)$ and $b_{i}(\tau)$ defined, the
grand canonical partition function can be written as
\begin{equation}
Z=\text{Tr}{e^{-\beta \hat{H}}}=\int \mathcal{D}b^{\ast}\mathcal{D}%
be^{-S[b^{\ast},b]/\hbar}\ ,\label{eq:eq21}%
\end{equation}
where $\mathcal{D}b^{\ast}$ and $\mathcal{D}b$ denote functional integration
for $b^{\ast}$ and $b$, respectively. The action $S[b^{\ast},b]$ is given by
\begin{align}
S[b^{\ast},b] &  =\int_{0}^{\hbar \beta}d\tau \Bigg[\sum_{i}b_{i}^{\ast}%
(\tau)\left(  \hbar \frac{\partial}{\partial \tau}-\mu \right)  b_{i}%
(\tau)\nonumber \\
&  -\sum_{ij}t_{ij}b_{i}^{\ast}(\tau)b_{j}(\tau)\nonumber \\
&  +\frac{1}{2}U\sum_{i}b_{i}^{\ast}(\tau)b_{i}^{\ast}(\tau)b_{i}(\tau
)b_{i}(\tau)\Bigg]\ ,
\end{align}
with $\beta=\frac{1}{k_{B}T}$, $k_{B}$ the Boltzmann constant, and $T$ the
temperature. With a Hubbard-Stratonovich transformation, we can rewrite the
action as \begin{widetext}
\begin{eqnarray}
S[b^*, b, \psi^*, \psi] &=& S[b^*, b] + \int_0^{\hbar \beta} d\tau \sum_{ij}\big( \psi_i^*(\tau) -b_i^*(\tau) \big)t_{ij}\big( \psi_j (\tau)-b_j(\tau) \big) \nonumber \\
&=& \int_0^{\hbar \beta} d\tau \Bigg[ \sum_i b_i^*(\tau) \left( \hbar \frac{\partial}{\partial \tau} -\mu \right)b_i(\tau)
+  \frac{1}{2}U\sum_i b_i^*(\tau)b_i^*(\tau)b_i(\tau)b_i(\tau) \nonumber \\
&&- \sum_{ij} \big(b_i^*(\tau)t_{ij} \psi_j(\tau) + \psi_i^*(\tau)t_{ij} b_j(\tau)\big)
+ \sum_{ij} \psi_i^*(\tau)t_{ij} \psi_j(\tau) \Bigg] \ ,
\label{eq:eq23}
\end{eqnarray}
\end{widetext}where $\psi^{\ast}$ and $\psi$ are the order parameter fields.
Hnece we have
\begin{align}
&  e^{-S^{\text{eff}}[\psi^{\ast},\psi]}=\exp{\left \{  -\frac{1}{\hbar}%
\int_{0}^{\hbar \beta}d\tau \sum_{ij}t_{ij}\psi_{i}^{\ast}(\tau)\psi_{j}%
(\tau)\right \}  }\nonumber \\
&  \times \int \mathcal{D}b^{\ast}\mathcal{D}b\exp \Bigg \{-S^{(0)}[b^{\ast
},b]/\hbar-\frac{1}{\hbar}\int_{0}^{\hbar \beta}d\tau \nonumber \\
&  \left(  -\sum_{ij}t_{ij}\big(b_{i}^{\ast}(\tau)\psi_{j}(\tau)+\psi
_{i}^{\ast}(\tau)b_{j}(\tau)\big)\right)  \Bigg \} \ ,\label{eq:eq24}%
\end{align}
where $S^{(0)}$ is the action with no hoping terms.

Using the cumulate expansion formula
\begin{equation}
\left \langle e^{A_{i}}\right \rangle =e^{\left \langle A_{i}\right \rangle
+\frac{1}{2}\left(  \left \langle A_{i}^{2}\right \rangle -\left \langle
A_{i}\right \rangle ^{2}\right)  +\ldots}\ , \label{eq:eq25}%
\end{equation}
we can get the effective action up to second-order
\begin{equation}
S^{\text{eff}}[\psi^{\ast},\psi]\approx S^{(0)}+S^{(2)}\ , \label{eq:eq26}%
\end{equation}
where perturbation term is \begin{widetext}
\begin{eqnarray}
S^{(2)}[\psi^*, \psi] &=& \int_0^{\hbar \beta} d\tau \sum_{ij} t_{ij} \psi_i^*(\tau)\psi_j(\tau)  -\frac{1}{2\hbar}\left< \left( \int_0^{\hbar \beta} d\tau \sum_{ij} t_{ij}\big[ b_i^*(\tau)\psi_j(\tau) + \psi_i^*(\tau)b_j(\tau) \big] \right)^2 \right>_{S^{(0)}} \nonumber \\
&=& \int_0^{\hbar \beta} d\tau \sum_{ij}t_{ij}\psi_i^*(\tau)\psi_j(\tau) \nonumber \\
&&- \frac{1}{2\hbar}\left<\int_0^{\hbar \beta}\int_0^{\hbar \beta}d\tau d\tau' \sum_{iji'j'} t_{ij}t_{i'j'}\big[ b_i^*(\tau)\psi_j(\tau) + \psi_i^*(\tau)b_j(\tau) \big]\big[ b_{i'}^*(\tau')\psi_{j'}(\tau') + \psi_{i'}^*(\tau')b_{j'}(\tau') \big]\right>_{S^{(0)}}\ .
\label{eq:eq27}
\end{eqnarray}
\end{widetext}
With the correlations in the unperturbed system
\begin{align}
&  \left \langle b_{i}^{\ast}b_{j}^{\ast}\right \rangle _{S^{(0)}}=\left \langle
b_{i}b_{j}\right \rangle _{S^{(0)}}=0\nonumber \\
&  \left \langle b_{i}^{\ast}b_{j}\right \rangle _{S^{(0)}}=\left \langle
b_{i}b_{j}^{\ast}\right \rangle _{S^{(0)}}=\left \langle b_{i}b_{i}^{\ast
}\right \rangle _{S^{(0)}}\delta_{ij}\ , \label{eq:eq28}%
\end{align}
we get
\begin{align}
&  S^{(2)}[\psi^{\ast},\psi]=\int_{0}^{\hbar \beta}d\tau \sum_{ij}\psi_{i}%
^{\ast}t_{ij}\psi_{j}-\frac{1}{\hbar}\int_{0}^{\hbar \beta}\int_{0}^{\hbar
\beta}d\tau d\tau^{\prime}\nonumber \\
&  \sum_{iji^{\prime}j^{\prime}}t_{ij}t_{i^{\prime}j^{\prime}}\psi_{j}^{\ast
}(\tau)\left \langle T_{\tau}[ b_{i}(\tau)b_{i^{\prime}}^{\ast}(\tau^{\prime
})]\right \rangle _{S^{(0)}}\psi_{j^{\prime}}(\tau^{\prime})\ . \label{eq:eq29}%
\end{align}

In general, the first term of above equation can be written as
\begin{equation}
\sum_{ij}t_{ij}\psi_{i}^{\ast}\psi_{j}=\sum_{\mathbf{k}}\bigg(\big(\psi
_{A\mathbf{k}}^{\ast}(\tau),\psi_{B\mathbf{k}}^{\ast}(\tau)\big)\mathcal{H}%
\Big(%
\begin{tabular}
[c]{c}%
$\psi_{A\mathbf{k}}(\tau)$\\
$\psi_{B\mathbf{k}}(\tau)$%
\end{tabular}
\  \Big)\bigg)\ ,\label{eq:eq30}%
\end{equation}
by a Fourier transformation, where $T_{\tau}$ is the imaginary-time order
operator. 
And the quadratic term becomes
\begin{align}
&  \sum_{jii^{\prime}j^{\prime}}t_{ji}t_{i^{\prime}j^{\prime}}\psi
_{\mathbf{r}_{j}}^{\ast}(\tau)\left \langle T_{\tau}[b_{\mathbf{r}_{i}}%
(\tau)b_{\mathbf{r}_{i^{\prime}}}^{\ast}(\tau^{\prime})]\right \rangle
_{S^{(0)}}\psi_{\mathbf{r}_{j^{\prime}}}(\tau^{\prime})\nonumber \\
&  =\sum_{jij^{\prime}}t_{ji}t_{ij^{\prime}}\psi_{\mathbf{r}_{j}}^{\ast}%
(\tau)\left \langle T_{\tau}[b_{\mathbf{r}_{i}}(\tau)b_{\mathbf{r}_{i}}^{\ast
}(\tau^{\prime})]\right \rangle _{S^{(0)}}\psi_{\mathbf{r}_{j^{\prime}}}%
(\tau^{\prime})\nonumber \\
&  =\left \langle T_{\tau}[b_{\mathbf{r}_{i}}(\tau)b_{\mathbf{r}_{i}}^{\ast
}(\tau^{\prime})]\right \rangle _{S^{(0)}}\sum_{jij^{\prime}}t_{ji}%
t_{ij^{\prime}}\psi_{\mathbf{r}_{j}}^{\ast}(\tau)\psi_{\mathbf{r}_{j^{\prime}%
}}(\tau^{\prime})\,.\label{eq:quad}%
\end{align}
It can be show that
\begin{align}
&  \sum_{jij^{\prime}}t_{ji}t_{ij^{\prime}}\psi_{\mathbf{r}_{j}}^{\ast}%
(\tau)\psi_{\mathbf{r}_{j}^{\prime}}(\tau^{\prime})\nonumber \\
&  =\sum_{\mathbf{k}}(\psi_{A\mathbf{k}}^{\ast}(\tau),\psi_{B\mathbf{k}}%
^{\ast}(\tau))\mathcal{H}^{2}\left(
\begin{tabular}
[c]{c}%
$\psi_{A\mathbf{k}}(\tau^{\prime})$\\
$\psi_{B\mathbf{k}}(\tau^{\prime})$%
\end{tabular}
\  \right)  \ .\label{eq:ewl}%
\end{align}
Hence at the transition point where $S^{(0)}$ vanishes, the effective action
can be written as
\begin{align}
S^{\text{eff}}\  &  =\int_{0}^{\hbar \beta}\sum_{\mathbf{k}}(\psi_{A\mathbf{k}%
}^{\ast},\psi_{B\mathbf{k}}^{\ast})\mathcal{H}\left(
\begin{tabular}
[c]{c}%
$\psi_{A\mathbf{k}}$\\
$\psi_{B\mathbf{k}}$%
\end{tabular}
\  \right)  -\frac{1}{\hbar}\int_{0}^{\hbar \beta}\int_{0}^{\hbar \beta}d\tau
d\tau^{\prime}\nonumber \label{eq:action}\\
&  \langle T_{\tau}[b(\tau)b^{\ast}(\tau^{\prime})]\rangle \sum_{\mathbf{k}%
}(\psi_{A\mathbf{k}}^{\ast}(\tau),\psi_{B\mathbf{k}}^{\ast}(\tau
))\mathcal{H}^{2}\left(
\begin{tabular}
[c]{c}%
$\psi_{A\mathbf{k}}(\tau^{\prime})$\\
$\psi_{B\mathbf{k}}(\tau^{\prime})$%
\end{tabular}
\  \right)  \ .
\end{align}
This is consistent with the results of a two-site model in Ref.\cite{Lim}.

\subsection{Dispersion of collective modes}

After obtaining the effective action of this bosonic checkerboard model of
flat-band, we calculate the critical point of MI-SF transition and then the
dispersion of collective modes.

For the bosonic checkerboard model of flat-band, the hoping terms are given
by
\begin{equation}
t_{ij}=t_{ji}^{\ast}=\left \{
\begin{tabular}
[c]{cc}%
$te^{i\varphi}$ & \mbox{for NN terms, }\\
$\pm t^{\prime}$ & \mbox{for NNN terms, }\\
$t^{\prime \prime}$ & \mbox{for NNNN terms, }\\
$0$ & \mbox{otherwise.}
\end{tabular}
\  \  \right.  . \label{hopping}%
\end{equation}
Then the $\mathcal{H}$ matrix was obtained as\cite{Sun}
\begin{align}
\mathcal{H}=  &  \left[
\begin{tabular}
[c]{cc}%
$\mathcal{A}$ & $\mathcal{C}^{\ast}$\\
$\mathcal{C}$ & $\mathcal{B}$%
\end{tabular}
\  \right] \nonumber \\
=  &  4t^{\prime \prime}\cos k_{x}\cos k_{y}I+4t\cos \varphi(\cos \frac{k_{x}}%
{2}\cos \frac{k_{y}}{2})\sigma_{x}\nonumber \\
&  +4t\sin \varphi(\sin \frac{k_{x}}{2}\sin \frac{k_{y}}{2})\sigma_{y}%
+2t^{\prime}(\cos k_{x}-\cos k_{y})\sigma_{z}\ ,
\end{align}
where $I$ and $\sigma_{x}$, $\sigma_{y}$, $\sigma_{z}$ is the identity and
Pauli matrices. Hence
\begin{align}
\mathcal{A}  &  =2t^{\prime}(\cos k_{x}-\cos k_{y})+4t^{\prime \prime}\cos
k_{x}\cos k_{y}\ ,\\
\mathcal{B}  &  =-2t^{\prime}(\cos k_{x}-\cos k_{y})+4t^{\prime \prime}\cos
k_{x}\cos k_{y}\ ,\\
\mathcal{C}  &  =4t\cos \varphi \cos \frac{k_{x}}{2}\cos \frac{k_{y}}{2}%
+4it\sin \varphi \sin \frac{k_{x}}{2}\sin \frac{k_{y}}{2}\ . \label{eq:hmatrix}%
\end{align}

The $\left \langle T_{\tau}[b(\tau)b^{\ast}(\tau^{\prime})]\right \rangle
_{S^{(0)}}$ is just the negative of Matsubara Green's function and it can be
evaluated as
\begin{align}
\left \langle T_{\tau}[b(\tau)b^{\ast}(\tau^{\prime})]\right \rangle _{S^{(0)}}
&  =\theta(\tau-\tau^{\prime})(n+1)e^{-(-\mu+nU)(\tau-\tau^{\prime})/\hbar
}\nonumber \label{eq:eq33}\\
&  +\theta(\tau^{\prime}-\tau)ne^{(\mu-(n-1)U)(\tau^{\prime}-\tau)/\hbar}\ .
\end{align}
After Fourier transformation to Hubbard-Stratonovich fields in the Matsubara
frequencies,
\begin{align}
\psi_{A\mathbf{k}}(\tau)  &  =\frac{1}{\sqrt{\hbar \beta}}\sum_{m}%
e^{-i\omega \tau}\psi_{A\mathbf{k,\omega_{m}}}\nonumber \\
\psi_{B\mathbf{k}}(\tau)  &  =\frac{1}{\sqrt{\hbar \beta}}\sum_{m}%
e^{-i\omega \tau}\psi_{B\mathbf{k,\omega_{m}}}\ , \label{eq:eq34}%
\end{align}
the effective action is given by
\begin{align}
&  S^{\text{eff}}[\psi^{\ast},\psi]\nonumber \label{eq:eq35}\\
&  =\sum_{\mathbf{k},m}(\psi_{A\mathbf{k},m}^{\ast},\psi_{B\mathbf{k},m}%
^{\ast})\left(  \mathcal{H}-\mathcal{H}^{2}f_{\omega_{m}}\right)  (%
\begin{tabular}
[c]{c}%
$\psi_{B\mathbf{k},\omega_{m}}$\\
$\psi_{B\mathbf{k},\omega_{m}}$%
\end{tabular}
\  \ )\nonumber \\
&  =\sum_{\mathbf{k},m}(\psi_{A\mathbf{k},m}^{\ast},\psi_{B\mathbf{k},m}%
^{\ast})(-\hbar \mathbf{G}^{-1}(\mathbf{k},i\omega_{m}))(%
\begin{tabular}
[c]{c}%
$\psi_{A\mathbf{k},m}$\\
$\psi_{B\mathbf{k},m}$%
\end{tabular}
\  \ )\ ,
\end{align}
where
\begin{equation}
f_{\omega_{m}}=\frac{n+1}{-i\hbar \omega_{m}-\mu+nU}+\frac{n}{i\hbar \omega
_{m}+\mu-(n-1)U}\ , \label{eq:eq36}%
\end{equation}
and
\begin{align}
&  \hbar \mathbf{G}^{-1}(\mathbf{k},i\omega_{m})\nonumber \label{eq:eq37}\\
&  =\left(
\begin{tabular}
[c]{cc}%
$(\mathcal{A}^{2}+\mathcal{C}^{\ast}\mathcal{C})f_{\omega_{m}}-\mathcal{A}$
\thinspace & $(\mathcal{A}+\mathcal{B})\mathcal{C}^{\ast}f_{\omega_{m}%
}-\mathcal{C}^{\ast}$\\
$(\mathcal{A}+\mathcal{B})\mathcal{C}f_{\omega_{m}}-\mathcal{C}$ \thinspace &
$(\mathcal{B}^{2}+\mathcal{C}^{\ast}\mathcal{C})f_{\omega_{m}}-\mathcal{B}$%
\end{tabular}
\  \  \  \right)  \ .
\end{align}
By substituting $i\omega_{m}\rightarrow \omega_{m}$, we get the equation for
real energy $\det[\mathbf{G}^{-1}]=0$, namely
\begin{equation}
(\mathcal{A}\mathcal{B}-\mathcal{C}^{\ast}\mathcal{C})\left[  (\mathcal{A}%
\mathcal{B}-\mathcal{C}^{\ast}\mathcal{C})f_{\omega_{m}}^{2}-(\mathcal{A}%
+\mathcal{B})f_{\omega_{m}}+1\right]  =0. \label{eq:eq38}%
\end{equation}

\begin{figure}[ptbh]
\begin{center}
\includegraphics[width=0.45\textwidth]{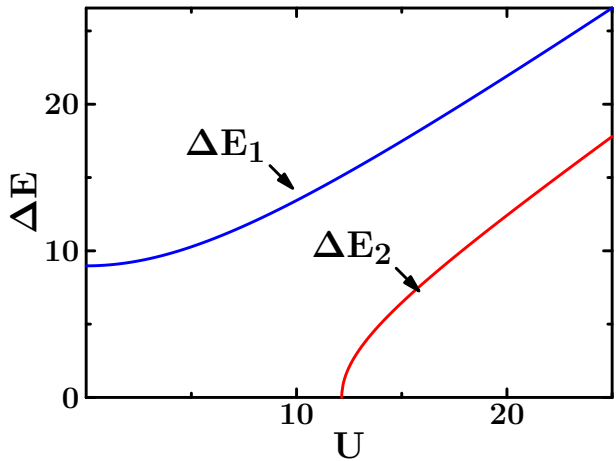}
\end{center}
\caption{The energy gaps of a pair of quasi-particle and quasi-hole, $\Delta
E_{1}$ and $\Delta E_{2}$ via $U$ when $n=1$ and $\mu=1$. One can see that MI-SF transition occurs at
$U=12.15t$. At MI-SF transition, $\Delta E_{1}$ is still finite while $\Delta
E_{2}$ close at the critical point. Hence the MI-SF transition is determined
by $\Delta E_{2}$.}%
\label{fig:E_U}%
\end{figure}

Solve Eq~\eqref{eq:eq36} and Eq~\eqref{eq:eq38}, we can get the
quasi-particle, quasi-hole spectra,
\begin{align}
\hbar \omega_{1,qp}  &  =\frac{1}{2}\left[  -2\mu+(2n-1)U-\frac{1}{f_{\omega
m+}^{2}}+\Delta E_{1,\mathbf{k}}\right]  \ ,\nonumber \\
\hbar \omega_{1,qh}  &  =\frac{1}{2}\left[  -2\mu+(2n-1)U-\frac{1}{f_{\omega
m+}^{2}}-\Delta E_{1,\mathbf{k}}\right]  \ ,\nonumber \\
\hbar \omega_{2,qp}  &  =\frac{1}{2}\left[  -2\mu+(2n-1)U-\frac{1}{f_{\omega
m-}^{2}}+\Delta E_{2,\mathbf{k}}\right]  \ ,\label{eq:eq39}\\
\hbar \omega_{2,qh}  &  =\frac{1}{2}\left[  -2\mu+(2n-1)U-\frac{1}{f_{\omega
m-}^{2}}-\Delta E_{2,\mathbf{k}}\right]  \ ,\nonumber
\end{align}
where
\begin{align}
\Delta E_{1,\mathbf{k}}  &  =\hbar \omega_{1,qp}-\hbar \omega_{1,qh}=\sqrt
{U^{2}-\frac{(4n+2)U}{f_{\omega m+}}+\frac{1}{f_{\omega m+}^{2}}}\ ,\\
\Delta E_{2,\mathbf{k}}  &  =\hbar \omega_{2,qp}-\hbar \omega_{2,qh}=\sqrt
{U^{2}-\frac{(4n+2)U}{f_{\omega m-}}+\frac{1}{f_{\omega m-}^{2}}}\ ,\\
f_{\omega m\pm}  &  =\frac{\mathcal{A}+\mathcal{B}\pm \sqrt{(\mathcal{A}%
-\mathcal{B})^{2}+4\mathcal{C}^{\ast}\mathcal{C}}}{2(\mathcal{A}%
\mathcal{B}-\mathcal{C}^{\ast}\mathcal{C})}\ . \label{eq:eq40}%
\end{align}
Hence $\Delta E_{1/2,\mathbf{k}}$ are the dispersion of elementary excitations
- a pair of quasi-particle and quasi-hole.

\begin{figure}[ptbh]
\begin{minipage}[c]{0.8\linewidth}
\includegraphics[width=\textwidth]{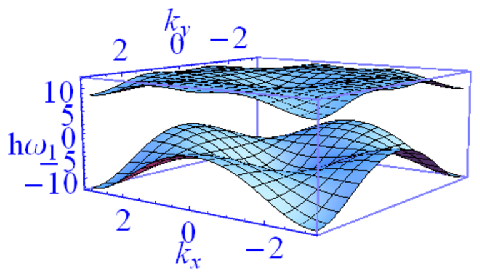}
\end{minipage}\newline \begin{minipage}[c]{0.7\linewidth}
\includegraphics[width=\textwidth]{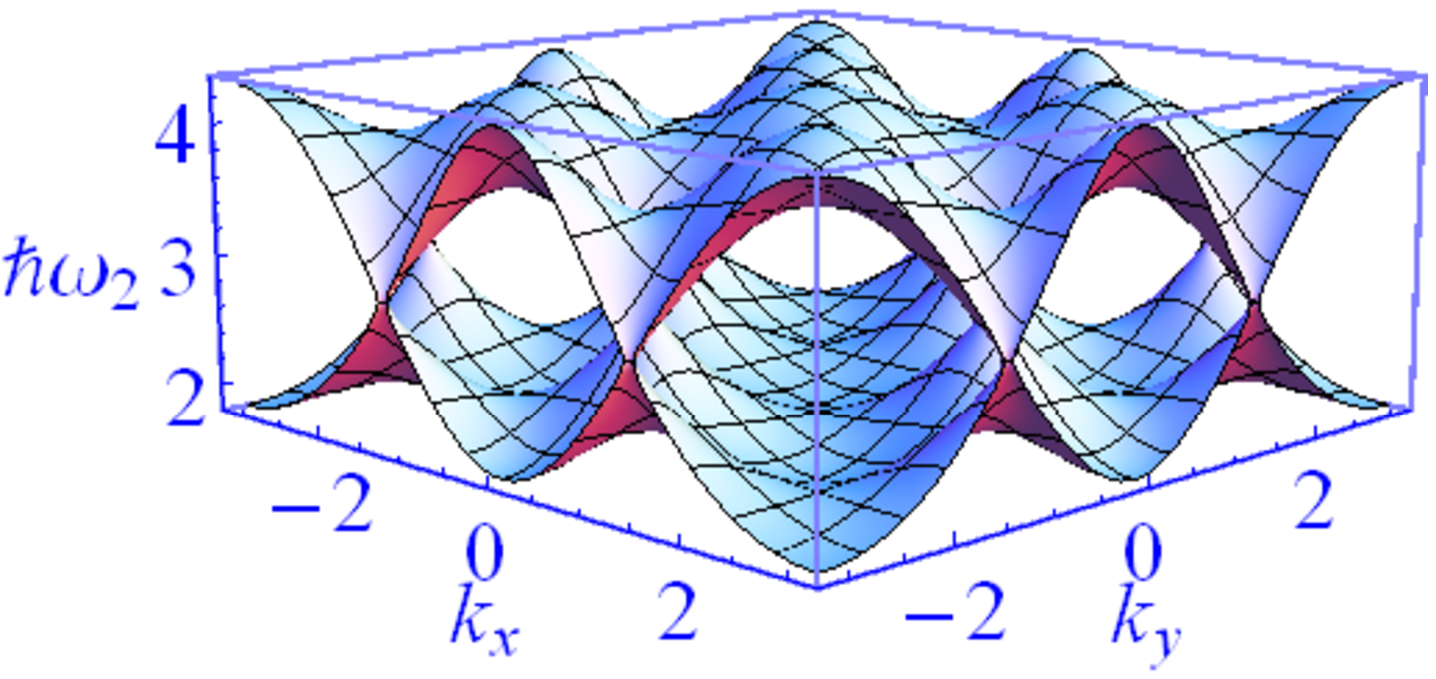}
\end{minipage}
\caption{The spectra of quasi-particle and quasi-hole for the case of critical
point, $U\approx12.15\left \vert t\right \vert $ when $n=1$ and $\mu=1$. The upper one is for
$\hbar \omega_{1,qp,qh}$ while the lower one is for $\hbar \omega_{2,qp,qh}$.}%
\label{fig:spectra1}%
\end{figure}

\begin{figure}[ptbh]
\begin{center}
\includegraphics[width=0.45\textwidth]{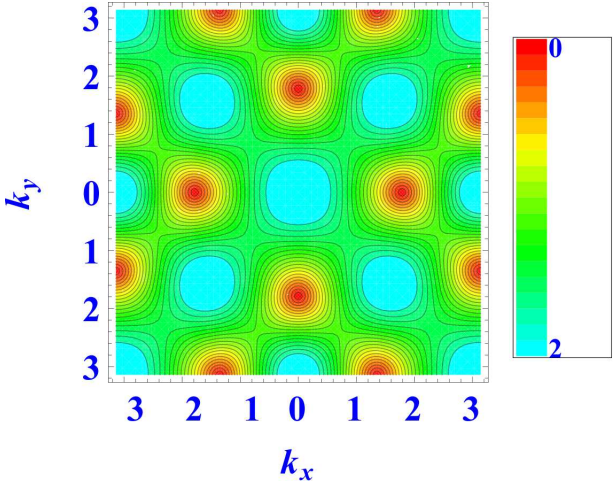}
\end{center}
\caption{The dispersion of a pair of quasi-particle and quasi-hole $\Delta
E_{2,\mathbf{k}}$ at the critical point with $U\approx12.15\left \vert
t\right \vert $ when $n=1$ and $\mu=1$.}%
\label{fig:dispersion1}%
\end{figure}

From the information of quasi-particle-quasi-hole spectra, we can determine
the phase boundary of MI-SF transition. We plot the energy gaps via $U$ in
Fig\ref{fig:E_U} with suitable $\mathbf{k}$ where the energy gaps take their
maximum value, $\Delta E_{1/2}=\hbar \omega_{1/2,qp}-\hbar \omega_{1/2,qh}$. The
MI-SF transition occurs at $U=U_{c}\approx12.15\left \vert t\right \vert $ when
the energy gap vanishes, which implies that $\Delta=\Delta E_{2}=0$. From the
results, we found that the topologically nontrivial flat-band only lightly
changes the phase boundary between superfluid phase and Mott phase.

In Fig.\ref{fig:spectra1}, we plot the spectra of two branches of
quasi-particle and quasi-hole, $\hbar \omega_{1,qp,qh}$ and $\hbar
\omega_{2,qp,qh}.$ One can see that in MI phase, the dispersion of a pair of
quasi-particle and quasi-hole $\Delta E_{1,\mathbf{k}}=\hbar \omega
_{1,qp}-\hbar \omega_{1,qh}$ are always have larger energy than $\Delta
E_{2,\mathbf{k}}=\hbar \omega_{2,qp}-\hbar \omega_{2,qh}$. Thus we focus on the
quasi-particle-quasi-hole excitations with dispersion $\Delta E_{2,\mathbf{k}%
}$ in the following parts. For this case ($U=U_{c}\approx12.15\left \vert
t\right \vert $), the dispersion relation of the two quasi-particle quasi-hole
spectra $\Delta E_{2,\mathbf{k}}$ is displayed in Fig.\ref{fig:dispersion1}.
At MI-SF transition we found that the dispersion of quasi-particles and
quasi-holes $\Delta E_{2,\mathbf{k}}$ show nodal-like behavior near special
points in momentum space at $\mathbf{k}=(0,\pm \pi/2)$, $\mathbf{k}=(\pm
\pi/2,0)$.

\begin{figure}[ptbh]
\begin{minipage}[c]{0.75\linewidth}
\includegraphics[width=\textwidth]{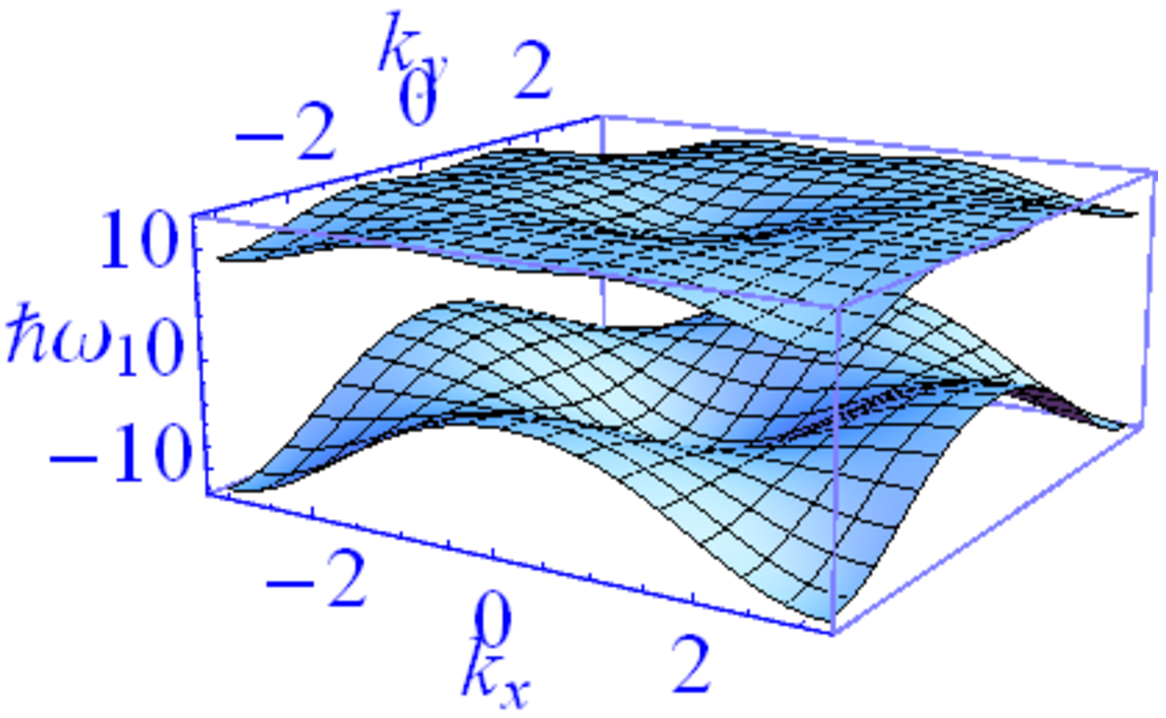}
\end{minipage}\newline \begin{minipage}[c]{0.8\linewidth}
\includegraphics[width=\textwidth]{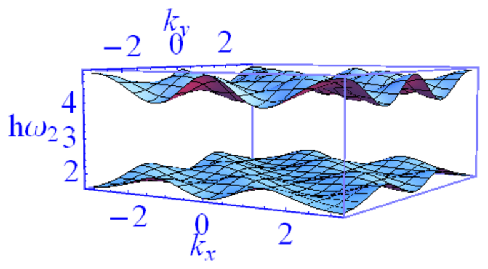}
\end{minipage}
\caption{The spectra of quasi-particle and quasi-hole for the case of
$U=12.5\left \vert t\right \vert $ when $n=1$ and $\mu=1$. The upper one is for $\hbar \omega_{1,qp,qh}$
while the lower one is for $\hbar \omega_{2,qp,qh}$.}%
\label{fig:spectra2}%
\end{figure}

\begin{figure}[ptbh]
\begin{center}
\includegraphics[width=0.45\textwidth]{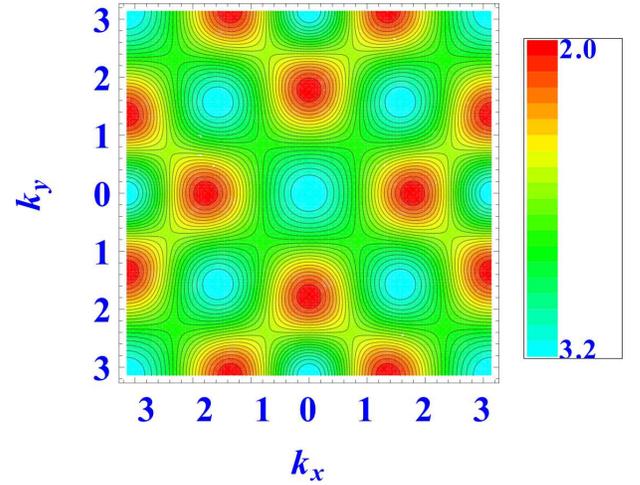}
\end{center}
\caption{The dispersion of the quasi-particle quasi-hole $\Delta
E_{2,\mathbf{k}}$ for $U=12.5\left \vert t\right \vert $ when $n=1$ and $\mu=1$. }%
\label{fig:dispersion2}%
\end{figure}

In MI region, near MI-SF transition we also plot the spectra of two branches
of quasi-particle and quasi-hole, $\hbar \omega_{1,qp,qh}$ and $\hbar
\omega_{2,qp,qh}$ for the case of $U=12.5\left \vert t\right \vert $ in
Fig.\ref{fig:spectra2}. In addition, the dispersion of a pair of
quasi-particle and quasi-hole is shown in Fig.\ref{fig:dispersion2}, from
which we found that the quasi-particle-quasi-hole excitations have energy gap
and the dispersion of a pair of quasi-particle and quasi-hole $\Delta
E_{2,\mathbf{k}}$ becomes flat. To illustrate this effect, we calculate the
flatness ratio, $\rho=\Delta/W$ where $W$ is the band width of the energy
spectra and $\Delta$ is the energy gap of of the excitations. The flatness ratio $\rho$ for $U=12.5\left \vert t\right \vert $ is about $1.8$.

\begin{figure}[ptbh]
\begin{minipage}[c]{0.8\linewidth}
\includegraphics[width=\textwidth]{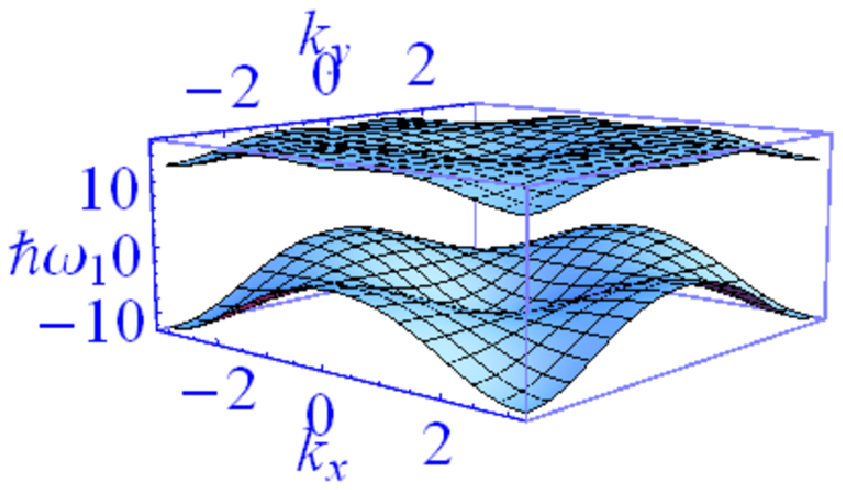}
\end{minipage}\newline \begin{minipage}[c]{0.8\linewidth}
\includegraphics[width=\textwidth]{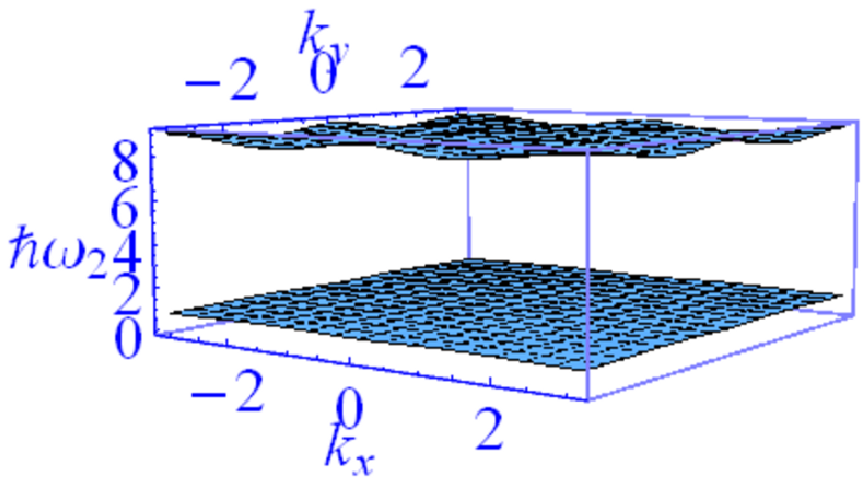}
\end{minipage}
\caption{The spectra of quasi-particle and quasi-hole for the case of
$U=16\left \vert t\right \vert $ when $n=1$ and $\mu=1$. The upper one is for $\hbar \omega_{1,qp,qh}$
while the lower one is for $\hbar \omega_{2,qp,qh}$.}%
\label{fig:spectra3}%
\end{figure}

\begin{figure}[ptbh]
\begin{center}
\includegraphics[width=0.45\textwidth]{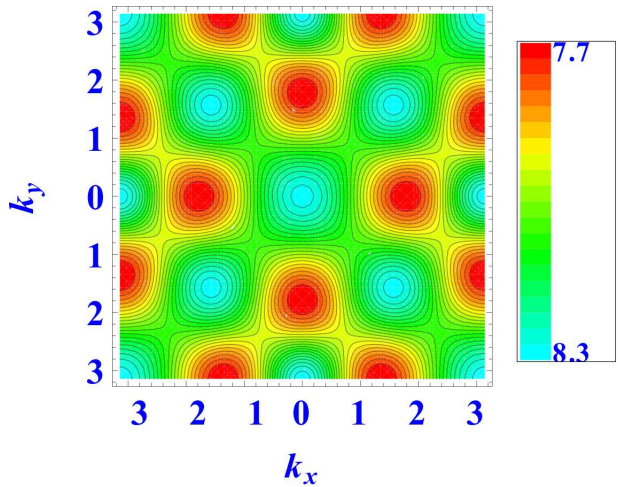}
\end{center}
\caption{The dispersion of a pair of quasi-particle and quasi-hole $\Delta
E_{2,\mathbf{k}}$ for $U=16\left \vert t\right \vert $ when $n=1$ and $\mu=1$. }%
\label{fig:dispersion3}%
\end{figure}

In MI region, far from MI-SF transition, we plot the spectra of two branches
of quasi-particle and quasi-hole $\hbar \omega_{1,qp,qh}$ and $\hbar
\omega_{2,qp,qh}$ and the dispersion for this case of $U=16\left \vert
t\right \vert $ in Fig.\ref{fig:spectra3} and Fig.\ref{fig:dispersion3},
respectively. In this figure, we can see that there exists a flat-band of
$\Delta E_{2,\mathbf{k}}$ obviously. Now the flatness ratio $\rho$ for
$U=12.5\left \vert t\right \vert $ is about $16$.

\begin{figure}[ptbh]
\begin{center}
\includegraphics[width=0.5\textwidth]{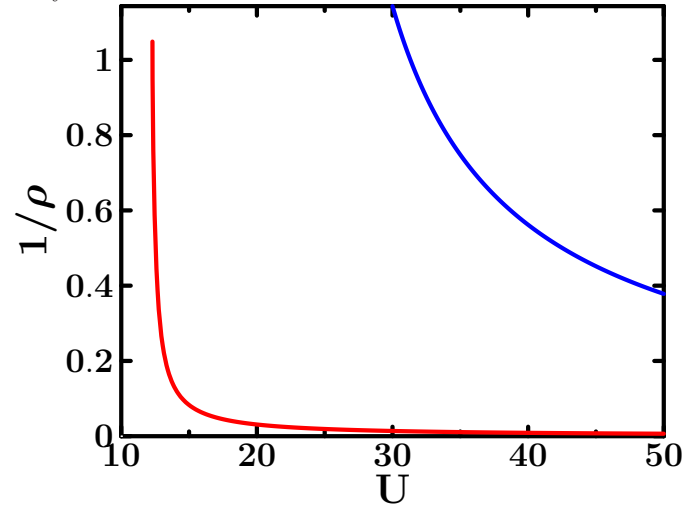}
\end{center}
\caption{The inverse of flatness ratio $1/\rho$ via $U/\left \vert t\right \vert $ when $n=1$ and $\mu=1$, where the
red line is that of flat-band model while blue line is that of traditional
Bose-Hubbard model. One can see that flatness ratio of the generalized
Bose-Hubbard on a checkerboard model with topologically nontrivial flat-band
increases much more rapidly, implying that there exists flat-band for the
bosonic excitations.}%
\label{fig:ratio}%
\end{figure}

Above results indicate that the dispersion of a pair of quasi-particle and
quasi-hole will become more and more flat when we increase interaction
strength, $U$. In Fig\ref{fig:ratio}, we displayed the inverse of flatness
ratio $\rho$ via $U$ of generalized Bose-Hubbard on a checkerboard model with
topologically nontrivial flat-band. From these results, we can see that the
flatness ratio of the generalized Bose-Hubbard on a checkerboard model with
topologically nontrivial flat-band increases with increasing of the interaction
strength. On the other hand, we also calculate the flatness ratio of
traditional Bose-Hubbard model. See blue line in Fig.\ref{fig:ratio}. From it
we can see that the flatness ratio of traditional Bose-Hubbard model changes
much more slowly than the flat-band model with increasing of the interaction strength. In this sense we can say that
in MI phase, there indeed exist flat bands for (bosonic) quasi-particle or
quasi-hole for the generalized Bose-Hubbard on a checkerboard model with
topologically nontrivial flat-band.

\section{Conclusion}

In this paper, using a decoupling approximation, we studied the generalized
Bose-Hubbard on a checkerboard model with topologically nontrivial flat-band
in the mean-field level. We find that the MI-SF phase transition of the
flat-band case only lightly changes compared with the traditional Bose-Hubbard
model. We also calculate dispersion relations of the collective modes Mott
phase. The results show that in MI phase the (bosonic) quasi-particle or
quasi-hole also has flat bands. In the end, we should point out that until now
we have no idea about how to realize this bosonic checkerboard model of
flat-band in optical lattice of cold atoms. In the future we will revisit this
issue and find way to realize this bosonic checkerboard model of flat-band
with particular big nearest-neighbor (NN) and the next-nearest-neighbor (NNN) hoppings.

\begin{acknowledgments}
This work is supported by NFSC Grant No. 11174035, National Basic Research
Program of China (973 Program) under the grant No. 2011CB921803, 2012CB921704.
\end{acknowledgments}


\begin{thebibliography}{99}                                                                                               %
  \bibitem{Jaksch} D. Jaksch, C. Bruder, J.I. Cirac, C. W. Gardiner, and P. Zoller, Phys. Rev. Lett. \textbf{81}, 3108 (1998).

\bibitem {Greiner}M. Greiner, O. Mandel, T. Esslinger, T. W. H\"{a}nsch, and
I. Bloch, Nature \textbf{415}, 39 (2002).

\bibitem {Buluta}I. Buluta and F. Nori, Science \textbf{326}, 108 (2009).

\bibitem {Fisher}M. P. A. Fisher, P. B. Weichman, G. Grinstein, and D. S.
Fisher, Phys. Rev. B \textbf{40}, 546 (1989).

\bibitem {haldane}F. D. M. Haldane, Phys. Rev. Lett. \textbf{61}, 2015 (1988).

\bibitem {Tang}E. Tang, J.-W. Mei, and X.-G. Wen, Phys. Rev. Lett.
\textbf{106} 236802 (2011).

\bibitem {Sun}K. Sun, Z. C. Gu, H. Katsura, and S. Das Sarma, Phys. Rev. Lett.
\textbf{106}, 236803 (2011).

\bibitem {Neupert}T. Neupert, L. Santos, C. Chamon, and C. Murdy, Phys. Rev.
Lett. \textbf{106}, 236804 (2011).

\bibitem {Wang}Y.-F. Wang, Z.-C. Gu, C.-D Gong and D. N. Sheng, Phys. Rev.
Lett. \textbf{107}, 146803 (2011).

\bibitem {Oosten}D. van Oosten, P. van der Straten, and H. T. C. Stoof, Phys.
Rev. A \textbf{63}, 053601 (2001).

\bibitem {Lim}L.-K. Lim, A. Hemmerich, and C. M. Smith, Phys. Rev. A
\textbf{81}, 023404 (2010).
\end{thebibliography}
\end{document}